# Silicon Nitride Waveguides with Intrinsic Single-Photon Emitters for Integrated Quantum Photonics

Alexander Senichev[1,2], Samuel Peana[1,2], Zachariah O. Martin[1,2], Omer Yesilyurt[1,2],

Demid Sychev[1,2], Alexei S. Lagutchev[1,2], Alexandra Boltasseva[1,2], Vladimir M. Shalaev[*,1,2]

[1]Elmore Family School of Electrical and Computer Engineering, Birck Nanotechnology Center and Purdue Quantum Science and Engineering Institute, Purdue University, West Lafayette, IN 47907, USA

[2]Quantum Science Center, a National Quantum Information Science Research Center of the U.S. Department of Energy, Oak Ridge, TN 37931, USA

**ABSTRACT:** The recent discovery of room temperature intrinsic single-photon emitters in silicon nitride (SiN)[1] provides the unique opportunity for seamless monolithic integration of quantum light sources with the well-established SiN photonic platform. In this work, we develop a novel approach to realize planar waveguides made of low-autofluorescing SiN with intrinsic quantum emitters and demonstrate the single-photon emission coupling into the waveguide mode. The observed emission coupling from these emitters is found to be in line with numerical simulations. The coupling of the single-photon emission to a waveguide mode is confirmed by second-order autocorrelation measurements of light outcoupled off the photonic chip by grating couplers. Fitting the second-order autocorrelation histogram yields $g^{(2)}(0) = 0.35 \pm 0.12$ without spectral filtering or background correction with an outcoupled photon rate of $10^4$ counts per second. This demonstrates the first successful coupling of photons from intrinsic single-photon emitters in SiN to monolithically integrated waveguides made of the same material. The results of our work pave the way toward the realization of scalable, technology-ready quantum photonic integrated circuitry efficiently interfaced with solid-state quantum emitters.

## INTRODUCTION

Photonic quantum systems employ photons to encode and transmit quantum information and distribute quantum entanglement through interaction with other photons and various stationary qubits. Solid-state quantum emitters are fundamental resources for photon-based quantum information technologies. They can serve as on-demand single-photon sources, components of atomic memories, quantum repeaters, and quantum sensors[2-6]. Realization of these devices requires photonic elements, which would enable efficient interfacing between light and matter, high collection efficiency, low-loss routing of photons, and interconnection between photonic qubits for quantum logic operations[7,8]. The direct coupling of quantum emitters with quantum photonic integrated circuits (QPICs) offers scalability and stability to photonic elements. This in turn may enable a high density of on-chip photonic qubit sources and the level of performance required for practical applications in the quantum domain[8-10].

The possibility to integrate quantum emitters with QPICs critically depends on the properties of the host materials. The variety of well-established solid-state quantum emitters includes III-V semiconductor quantum dots (QDs)[11-13], defect-based color centers in: diamond[14-16], silicon carbide (SiC)[17-20], and hexagonal boron nitride (hBN)[21-23], to name a few. The high brightness, single-photon purity, indistinguishability, and optically addressable spins of these quantum emitters make them promising for applications in quantum communication, computing, and sensing. Some of these materials hosting quantum emitters can be used for the fabrication of photonic components for QPICs. For example, diamond QPICs with embedded quantum emitting color centers were recently used to demonstrate on-chip cavity-enhanced single-photon sources and quantum memories[5,24-26]. III-V semiconductors, SiC, and hBN have also been extensively studied as platforms with intrinsic quantum emitters which are also capable of supporting QPIC elements such as: waveguides, directional couplers, and resonant cavities among others[27-29]. However, these materials support the fabrication of a limited variety of photonic components insufficient for full-fledged QPICs. Hence, on-chip integration of common solid-state quantum emitters typically relies on hybrid approaches combining materials hosting quantum emitters and material platforms well-developed for QPICs[30,31].

Currently, most developed, technology-driven material platforms for integrated photonics include: silicon on insulator (SOI), silicon nitride (SiN), lithium niobate (LiNbO$_3$), and aluminum nitride (AlN)[8]. Large-scale hybrid integration was recently demonstrated for quantum emitters in diamond coupled to AlN photonic circuitry[32]. The semiconductor quantum dots were integrated with LiNbO$_3$ photonic devices[33]. Hybrid integration with SiN photonic circuitry has also been demonstrated for various quantum emitters such as: semiconductor quantum dots[34,35], color centers in 2D materials[36–38], and nanodiamonds[39,40]. Although the hybrid integration approach provides more flexibility in the realization of large-scale QPICs and has shown tremendous progress, it also faces challenges related to inevitable photon losses due to necessity of interfacing dissimilar materials, and to complex integration geometries, and fabrication procedures.

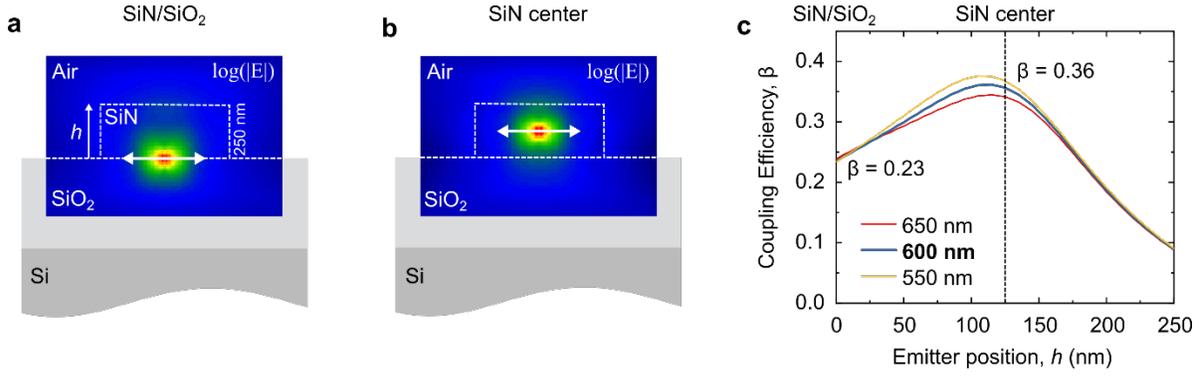

**Figure 1**. Coupling efficiency simulation of an intrinsic quantum emitter in a SiN waveguide. (a) Interface-positioned dipole with the simulated normalized electric field. (b) Center-positioned dipole and its normalized electric field. The dipole is oriented in the plane of the substrate and orthogonal to the waveguide axis for both cases is indicated with white two-headed arrows. (c) Simulated coupling efficiency as a function of the dipole position $h$ within the waveguide for three wavelengths 550 nm, 600 nm, and 650 nm.

Consequently, particular attention has been paid to engineering intrinsic or directly embedded quantum emitters in material platforms most suitable for the realization of QPICs. Such intrinsic quantum emitters have been discovered in gallium nitride[41,42], aluminum nitride[43–46], and silicon[47–49]. Monolithic integration of these intrinsic emitters with QPICs offers high coupling efficiencies along with simplified fabrication processes and it promises to enable the long-awaited industrial-scale quantum photonic circuitry.

Our group has recently discovered bright, stable, linearly polarized, and high-purity sources of single-photon emission in nitrogen-rich SiN operating at room temperature[1]. Stochiometric silicon nitride ($Si_3N_4$) is a leading integrated photonics material platform that is technologically mature, CMOS compatible and offers low propagation and insertion loss devices. It is well established for linear and nonlinear integrated optics[50–52] and has recently emerged as a potential platform for integrated quantum photonics[8,53,54]. Importantly, $Si_3N_4$ photonic components exhibit ultra-low losses, which is especially critical for the realization of large-scale QPICs. Recent examples of experimentally realized $Si_3N_4$-based QPICs elements include: on-chip frequency converters, phase shifters, and beam splitters all combined in a fully reconfigurable quantum photonic processor[53]. Consequently, $Si_3N_4$ is the material of choice for the development of the next generation of photonic quantum computers based on ultra-low loss waveguides by photonic quantum technology companies such as, for example, Xanadu[55] and QuiX[56].

Here we report on the development of low-autofluorescence SiN planar waveguides containing intrinsic single-photon sources recently discovered in our previous work[1]. While stoichiometric $Si_3N_4$ provides the ultra-low loss photonic waveguides, non-stoichiometric nitrogen-rich SiN used in our work offers substantially lower background fluorescence in the visible spectral range. This facilitates the integration with visible-wavelength quantum emitters[1,57]. In this work, we demonstrate addressing of the individual intrinsic single-photon sources in SiN photonic structures. To accomplish this, we grew low-autofluorescence SiN films with bright, stable, room-temperature quantum emitters on $SiO_2$ coated silicon substrates. From this film, we fabricated SiN waveguides with on-chip grating couplers. The photophysical properties of SiN single-photon emitters in the waveguides were characterized collecting photons both directly outcoupled to the far-field and coupled to the waveguide mode and then outcoupled off-chip by grating couplers. Our findings show that quantum emitters retain their optical properties after the waveguide fabrication procedure. With these results we demonstrate the first successful realization of low-autofluorescing SiN waveguides with intrinsic quantum emitters and the coupling of single-photon emission to a waveguide mode.

## RESULTS AND DISCUSSION

Utilizing the previously developed fabrication approach, we grow a material stack with low-autofluorescing SiN containing intrinsic quantum emitters[1]. This involves the growth of nitrogen-rich SiN films using high density plasma chemical vapor deposition (HDPCVD) on commercially available silicon dioxide ($SiO_2$) coated silicon substrates. Subsequently, the intrinsic single photon emitters in SiN are generated by post-growth rapid thermal annealing (RTA) for 120 seconds at a temperature of 1100°C in a nitrogen atmosphere. Ellipsometry characterization of the samples revealed that the refractive index $n$ of the resultant SiN films was ~1.7. This value is lower than the typical refractive index of $n$=2.0 reported for stoichiometric $Si_3N_4$, which is an acceptable trade-off at the current stage to obtain low-autofluorescing SiN[57,58].

The waveguides were designed and simulated with a commercial Maxwell solver. The dipole orientation for all simulations was taken in plane of the substrate and orthogonal to the waveguide axis. Such dipole orientation was selected as an optimal one for the coupling to the fundamental waveguide mode. However, in the experiment we do not control the dipole orientation. From the previous work, we expect the quantum emitters to be located at the interface between $SiO_2$ and SiN layers[1]. Simulation results for an interface-positioned quantum emitter (**Fig. 1a**) show coupling efficiency β=23% of the emission from the dipole source to the fundamental waveguide mode for the given SiN refractive index of 1.7 and the central wavelength of 600 nm. (**Fig. 1c**). It is important to note that this value for coupling efficiency β for the given material system is already better than those typically reported for the hybrid integration geometries[38,59]. For example, the simulated coupling efficiencies for an emitter placed on the top of a SiN waveguide[59] and embedded into a SiN waveguide[38] were found to be a few percent and up to approximately 19%, respectively. The near-optimal coupling efficiency to the fundamental waveguide mode for the given waveguide geometry can be obtained for the central position of the emitting dipole yielding β=36% (**Fig. 1b-c**). Moreover, the coupling efficiency of intrinsic SiN quantum emitters can be even further improved by precision positioning of an emitter into topology optimized couplers as has been shown theoretically[60]. This is the subject of the ongoing effort that will later be reported elsewhere. We also simulated the outcoupling efficiency from the grating couplers at 600 nm resulting in η=22%.

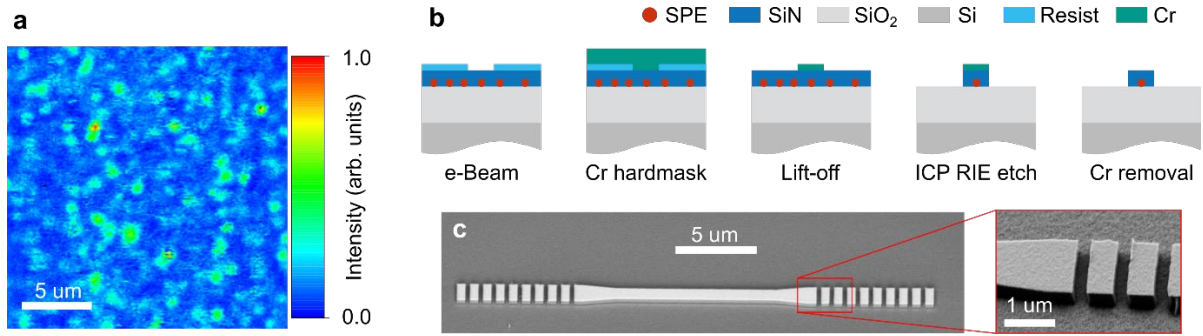

**Figure 2**. Fabrication of low-autofluorescence SiN waveguides with intrinsic quantum emitters. (a) 20x20 μm² confocal PL intensity map of the SiN sample before waveguide fabrication revealing a high density of isolated emitters. (b) The fabrication process used to create waveguides out of continuous SiN films containing intrinsic quantum emitters. (c) SEM images of a fabricated waveguide with an enlarged view of a grating coupler segment.

The photophysical properties of quantum emitters in SiN samples were characterized before waveguide fabrication using our custom-made scanning confocal microscope equipped with a Hanbury-Brown-Twiss (HBT) setup (see **Methods** for details). The observed emitters exhibit properties in line with those previously reported[1]. The confocal photoluminescence (PL) map shown in **Fig. 2a** reveals bright isolated emitters with an average density of approximately 10 emitters per 100 μm². The dimensions of the simulated waveguides and the high density of emitters ensures a high probability for each waveguide to contain at least several emitters without any spatial alignment procedure. A large array of waveguides was fabricated using conventional electron-beam lithography (EBL) and inductively coupled plasma reactive ion etching (ICP RIE). The fabrication approach is schematically depicted in **Fig. 2b** (see **Methods** for details). **Fig. 2c** shows scanning electron microscopy (SEM) images of a representative SiN waveguide before the Cr-mask removal.

After fabrication, the samples with waveguides were characterized with our confocal microscope addressing individual waveguide-coupled quantum emitters. **Figure 3a** shows a confocal PL intensity map of a waveguide segment. There are a few quantum emitters clearly resolved within this 10-μm-long segment which is outlined by the dashed lines. The number of observed emitters is in-line with their density in the original SiN films. These quantum emitters can be now excited by coupling laser light into the waveguide via the grating couplers. **Figure 3b** is a wide-field image of the waveguide taken with a charged coupled camera (CCD) showing transmission of the 532-nm laser pump light for the polarization orthogonal to the waveguide axis. The laser intensity coupling and outcoupling spots at the waveguide gratings for the given polarization orientation were found to be of comparable intensity (see more data on laser coupling in the **Supporting Information**). The laser light coupled to the waveguide excites quantum emitters as it propagates. The PL from quantum emitters partially outcouples to the far-field and can be observed with the CCD camera. **Figure 3c** shows emission from spatially isolated quantum emitters positioned along the waveguide. Additional CCD images of the waveguides with quantum emitters are provided in the **Supporting Information**, confirming appearance of at least several quantum emitters per waveguide using our alignment-free approach. The same quantum emitter is circled in red in both the confocal PL map (**Fig. 3a**) and the CCD image (**Fig. 3b**), and was selected for further analysis.

With the selected quantum emitter, we explored different excitation-detection schemes for addressing individual emitters embedded in a SiN waveguide. First, we probed the emitter in a conventional confocal measurement scheme, where the PL signal is collected directly from the excitation spot (**Fig. 4a**). The PL spectrum has similar structure to those of SiN quantum emitters at room temperature reported previously (**Fig. 4b**)[1].

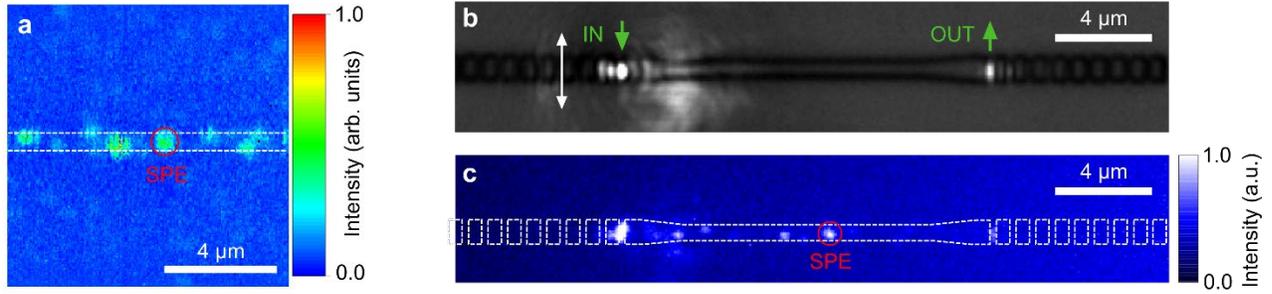

**Figure 3**. Spatial distribution of waveguide-integrated quantum emitters. (a) 10x10 μm² confocal PL intensity map of the SiN sample after the fabrication of a waveguide. The white dotted lines show the outline of the waveguide. (b) CCD image showing the laser light coupled into the waveguide through a grating coupler and outcoupled from the other end of it. (c) CCD image of the same waveguide as in (b) with excitation light suppressed using a 550-nm long pass filter. Fluorescence from individual single-photon emitters is clearly visible. The red circle indicates the quantum emitter selected for further analysis.

The PL spectrum can be fitted with four Gaussian curves with peak wavelengths of 576 nm, 600 nm, 624 nm, and 674 nm, which are within the spectral range typically observed for these emitters. The second-order autocorrelation $g^{(2)}(\tau)$ measurements reveal a clear dip below 0.5 at zero delay time, indicating that the selected emitter is likely a single-photon source. Fitting the experimental data provides a $g^{(2)}(0)$ value at zero delay time of 0.27±0.06 – without background correction or spectral filtering (**Fig. 4c**). This $g^{(2)}(0)$ value is within the expected range compared to previously reported results[1].

Next, we used a scheme with excitation and detection spots decoupled. For the selected emitter, we coupled the laser light to the waveguide through the grating coupler and collected the PL signal outcoupled to the far-field directly from the emitter as shown schematically in **Fig. 4d**. The PL spectrum matches that measured using the direct excitation scheme confirming that we addressed the same quantum emitter in both measurements (**Fig. 4e**). The PL intensity is found to be lower than for the direct excitation scheme. This is explained by the lower excitation power reaching the emitter through the waveguide. The fitting of the second-order autocorrelation histogram shown in **Fig. 4f** gives a $g^{(2)}(0)$ value comparable, within experimental uncertainty, to that measured using the direct excitation scheme.

Finally, we studied coupling of single-photon emission from a quantum emitter to the waveguide mode. For this purpose, we again used the decoupled excitation and detection scheme, but with the emitter excited directly and the emission coupled to a waveguide and collected remotely from one of the grating couplers as shown in **Fig. 5a**.

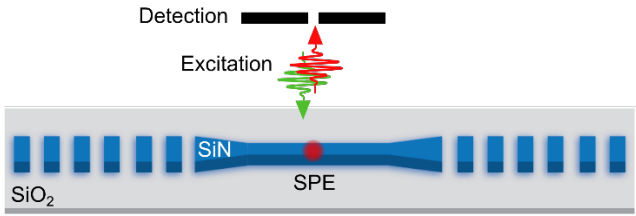
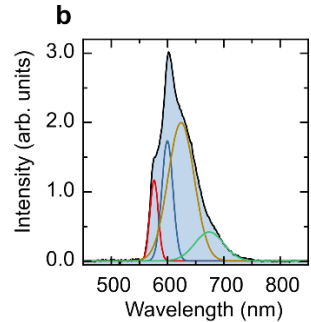
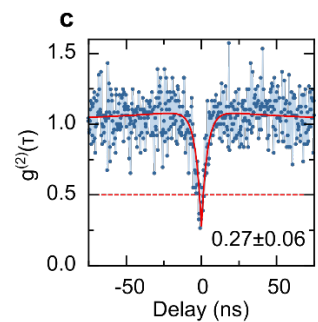
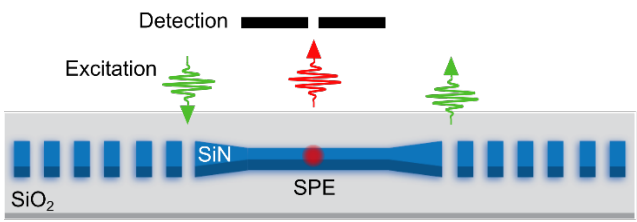
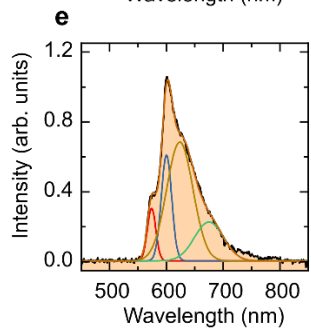
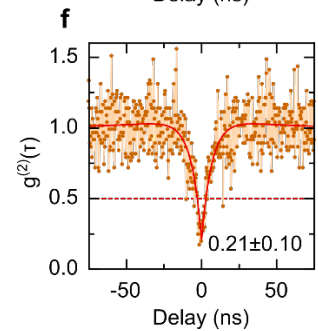

**Figure 4**. Photophysical properties of the selected quantum emitter measured in two configurations. (a) Direct excitation measurements of the quantum emitter showing its (b) PL spectrum and (c) second-order autocorrelation histogram $g^{(2)}(\tau)$. (d) Remote excitation measurements of the quantum emitter with the corresponding (e) PL spectrum and (f) second-order autocorrelation histogram $g^{(2)}(\tau)$. Both measurement schemes give comparable results.

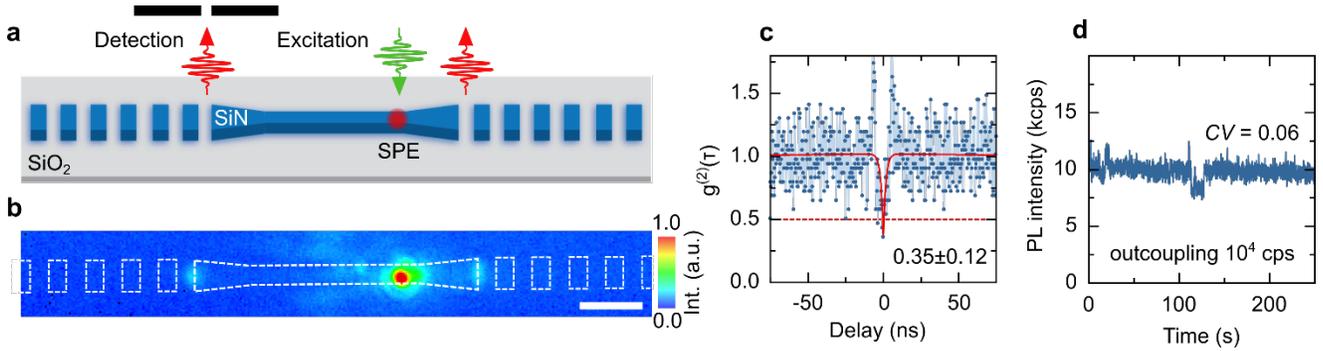

**Figure 5**. Observation of single-photon emission coupled to a waveguide mode. (a) Schematic of the decoupled excitation and detection scheme. (b) CCD image of the quantum emitter in the waveguide directly excited by the laser light. The excitation light is filtered out by a 550-nm long pass filter. Emission is collected from the left grating coupler: (c) Second-order autocorrelation histogram $g^{(2)}(\tau)$, (d) PL stability measurement showing minor blinking and low coefficient of variation (CV) for this particular emitter.

The direct excitation allows to address an emitter of choice avoiding the excitation of other emitters present in the waveguide. This method has previously been successfully used in similar configurations[29,45,59]. The laser polarization was aligned along the waveguide axis to reduce coupling into the waveguide, which can also excite other emitters and increase the unwanted background fluorescence.

**Figure 5b** shows the CCD image of the PL emission after the 550 nm long-pass filter is used to remove the 532-nm excitation light. The waveguide outline is shown with the dashed white lines. The excitation spot appears particularly bright due to the high-power laser used to excite the emitter and emission partially outcoupled into the far-field. The emission outcoupled at the grating couplers is also clearly visible in the CCD image. We performed the second-order autocorrelation $g^{(2)}(\tau)$ measurements using photons collected at one of the grating couplers to confirm the single-photon nature of the outcoupled emission. The $g^{(2)}(\tau)$ histogram is shown in **Fig. 5c** was collected over the course of 20 minutes to obtain a sufficient signal-to-noise ratio. The $g^{(2)}(0)$ value at a zero-delay time shows a pronounced dip below 0.5 indicating the outcoupled light still has single photon statistics. The value obtained from fitting the experimental data, $g^{(2)}(0) = 0.35 \pm 0.12$, is higher than that observed from the direct excitation-detection scheme. However, this decreased single-photon emission purity can most likely be explained by background noise from the SiN coupling to the waveguide mode or the presence of nearby emitters within the excitation area. The outcoupled emission rate was on the order of $10^4$ cps compared to the typical $10^6$ cps for the direct excitation-detection configuration. This indicates that the emitted photons were outcoupled through the waveguide as would be expected considering the simulated values for β and η. The emission coupling to the fundamental waveguide mode can be improved with the application of topology optimized couplers[60], while the collection efficiency can be substantially increased by using fiber end-fire coupling, which can provide theoretically up to 86% efficiency[61].

## CONCLUSION

In this work, we demonstrated first-time realization of low-autofluorescence SiN waveguides with intrinsic single-photon emitters that preserve their photophysical characteristics after all of the processing steps. The developed SiN waveguides utilize technology-ready material platform and feature native quantum emitters thus promising to enable much anticipated realization of scalable QPICs with a seamless integration of single photon sources. Notably, the discovered emitters in SiN provide a practical alternative to the previously reported complex, multi-materials integration schemes of single photon emitters with the SiN platform. The proposed integrated quantum emitters in SiN offer a great potential to enable a low-loss and scalable quantum photonic platform that is mature in terms of planar fabrication, quality control, and integration. Our findings call for further studies of SiN quantum emitters aimed at gaining a deeper understanding of their nature, properties control, devising schemes for site-controlled fabrication, and their scalable integration with on-chip quantum photonic circuitry.

## METHODS

**Sample growth and emitters activation.** The samples with SiN quantum emitters were prepared following the fabrication procedure developed in our previous work[1]. The SiN films were grown by a high-density plasma chemical vapor deposition (HDPCVD) system (Plasma-Therm Apex SLR) on commercially available silicon substrates coated with 3-um-thick $SiO_2$ (Rogue Valley Microdevices) making them suitable for waveguide fabrication. We used the nitrogen-rich growth conditions with a ratio of silicon to nitrogen precursors $N_2/SiH_4$ of 1.74 resulting in non-stoichiometric SiN. As was found in our previous work, these growth conditions give SiN films with low background autofluorescence in the visible spectral range similar to nitrogen-rich SiN reported in the literature[57]. Single-photon emitters were generated by post-growth rapid thermal annealing (RTA) for 120 seconds at a temperature of 1100°C in a nitrogen atmosphere using a Jipelec Jetfirst RTA system. The described procedure was found to provide reproducible fabrication of low auto-fluorescing SiN films with bright, stable, linearly polarized quantum emitters operating at room temperature.

**Waveguide design and fabrication.** All waveguides and outcouplers were designed and simulated with a commercial Maxwell solver (Lumerical) using the 3D finite-difference time-domain (FDTD) method. The waveguide dimensions were found to be 250 nm tall by 600 nm wide such that it supports the fundamental waveguide mode at the given emission wavelength. These dimensions were found assuming a refractive index 1.7 and the central wavelength of 600 nm. The quantum emitter is modeled with an electric dipole polarized in plane of the substrate perpendicular to the propagation axis. Since the dipole is not in a homogenous medium, the simulated coupling efficiencies β were normalized with corresponding Purcell factor. Outcoupled field was calculated from a flux through a horizontal plane placed 1 μm above the grating outcoupler. Outcoupling efficiency η is defined as the ratio of Poynting vector flux through this plane to the fundamental waveguide mode energy.

A large array of waveguides was fabricated on a single substrate to increase the odds of creating a waveguide with an optimally oriented emitter for efficient coupling. The design of waveguides was patterned into an electron beam resist (ZEP520A) via electron beam lithography and etched using inductively coupled plasma (ICP) reactive ion etching (RIE). Following development of the resist layer, a 20nm chrome film was evaporated onto the sample and lifted off to form a hardmask. The silicon nitride layer was then etched by ICP RIE. The chrome hard mask was then chemically removed using chrome etch (KMG CR-16). The high density of quantum emitters in the original SiN ensured the integration of at least few emitters in each waveguide without any special alignment procedures.

**Photoluminescence measurements.** The optical characterization of SiN waveguide-integrated quantum emitters was performed at room temperature. We used a custom-made scanning confocal microscope based on a commercial inverted microscope body (Ti–U, Nikon). The microscope was equipped with a 100 μm pinhole and a 100x air objective with a numerical aperture of 0.90 (Nikon). We estimate that the laser pump spot size on the sample was around 1 μm. The confocal scanning was performed with the objective mounted on a piezo stage (P-561, Physik Instrumente) driven by a controller (E-712, Physik Instrumente) and interfaced with LabVIEW (National Instruments). We used a 200-mW continuous wave 532 nm diode-pumped solid-state laser (Lambda beam PB 532-200 DPSS, RGB Photonics) for the optical excitation of emitters. The excitation light and photoluminescence (PL) signal were uncoupled by a 550 nm long-pass dichroic mirror (DMLP550, Thorlabs). The remaining pump power was further suppressed by a 550 nm long-pass filter (FEL0550, Thorlabs) installed in front of detectors. We acquired the emission with an avalanche detector with 69% quantum efficiency at 650 nm (SPCM-AQRH, Excelitas) for single-photon detection during scanning. To reveal the quantum nature of the emitters, second order autocorrelation function $g^{(2)}(\tau)$ measurements were performed using a Hanbury-Brown and Twiss (HBT) setup comprised of two avalanche detectors with a 30 ps time resolution and 35% quantum efficiency at 650 nm (PDM, Micro-Photon Devices), and an acquisition card with 4 ps internal jitter (SPC-150, Becker & Hickl). For the excitation of an emitter through the waveguide, we walked the laser spot along the waveguide using two mirrors in front of the microscope. The detection spot was kept at a grating coupler or a region of interest along the waveguide.

## ASSOCIATED CONTENT

**Supporting Information**. Polarization dependent laser coupling to a waveguide, additional CCD images of single-photon emitters integrated in SiN waveguides. This material is available free of charge via the Internet at http://pubs.acs.org.

## AUTHOR INFORMATION

### Corresponding Author

* E-mail: shalaev@purdue.edu.

### Author Contributions

The manuscript was written through contributions of all authors. All authors have given approval to the final version of the manuscript.

### Funding Sources

This work was supported by the U.S. Department of Energy (DOE), Office of Science through the Quantum Science Center (QSC), a National Quantum Information Science Research Center, National Science Foundation (NSF) grant 2015025-ECCS, and Purdue's Elmore ECE Emerging Frontiers Center "The Crossroads of Quantum and AI."

### Notes

The authors declare no competing financial interest.

# Supporting Information

# Silicon Nitride Waveguides with Intrinsic Single-Photon Emitters
# for Integrated Quantum Photonics

Alexander Senichev[1,2], Samuel Peana[1,2], Zachariah O. Martin[1,2], Omer Yesilyurt[1,2], Demid Sychev[1,2], Alexei S. Lagutchev[1,2], Alexandra Boltasseva[1,2], Vladimir M. Shalaev[*,1,2]

[1]Elmore Family School of Electrical and Computer Engineering, Birck Nanotechnology Center, Purdue University, and Purdue Quantum Science and Engineering Institute, West Lafayette, IN 47907, USA
[2]Quantum Science Center, a National Quantum Information Science Research Center of the U.S. Department of Energy, Oak Ridge, TN 37931, USA
[*]Email: shalaev@purdue.edu

1. **Laser Coupling**

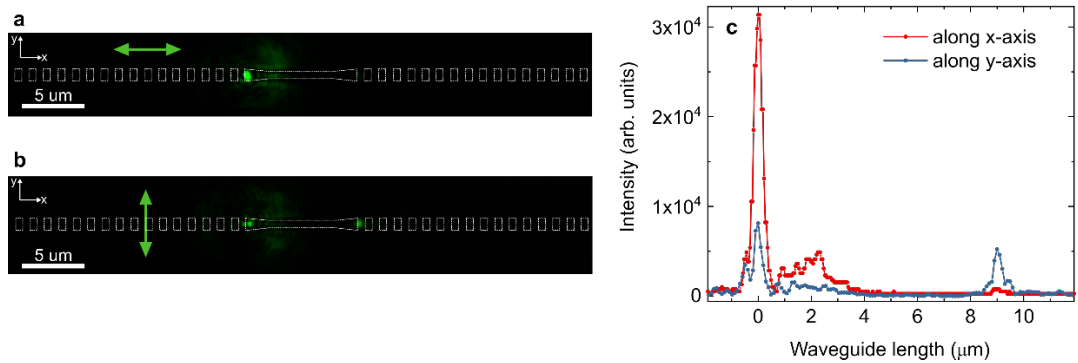

**Figure S1**. Coupling of the 532-nm pump laser light as a function of polarization. (a) Laser coupling with polarization along (x-axis) and (b) perpendicular (y-axis) to the waveguide. (c) Laser intensity profile along the waveguide for two different polarization orientations.

2. **CCD images of waveguide-integrated SiN quantum emitters**

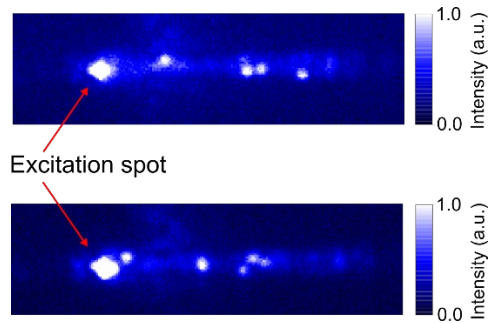

**Figure S2**. CCD images of two SiN waveguides showing fluorescence from individual single-photon emitters. The 550 nm long-pass filter in front of the CCD camera was used to filter out the 532 nm pump laser light.